\documentclass[conference]{IEEEtran}
\ifCLASSINFOpdf
\else
\fi

\usepackage{graphicx} %package to manage images
\graphicspath{ {./images/} }

\usepackage[rightcaption]{sidecap}

\usepackage{wrapfig}

\begin{document}
% Titles are generally ca  pitalized except for words such as a, an, and, as,
\title{United States Road Accident Prediction using Machine Learning Algorithms}



\author{\IEEEauthorblockN{Dominic Parosh Yamarthi}
\IEEEauthorblockA{Computer Science and Engineering \\
University at Buffalo\\
Email: dyamarth@buffalo.edu}
\and
\IEEEauthorblockN{Haripriya Raman}
\IEEEauthorblockA{Computer Science and Engineering \\
University at Buffalo\\
Email: hraman@buffalo.edu}
\and
\IEEEauthorblockN{Shamsad Parvin.PhD}
\IEEEauthorblockA{Computer Science and Engineering \\
University at Buffalo\\
Email: shamsadp@buffalo.edu}
}


%Telephone: (800) 555--1212, Fax: (888) 555--1212}


\maketitle

\begin{abstract}
Road accidents significantly threaten public safety and require in-depth analysis for effective prevention and mitigation strategies. This paper focuses on predicting accidents through the examination of a comprehensive traffic dataset covering 49 states in the United States. The dataset integrates information from diverse sources, including transportation departments, law enforcement, and traffic sensors. This paper specifically emphasizes predicting the number of accidents, utilizing advanced machine learning models such as regression analysis and time series analysis. The inclusion of various factors, ranging from environmental conditions to human behavior and infrastructure, ensures a holistic understanding of the dynamics influencing road safety. Temporal and spatial analysis further allows for the identification of trends, seasonal variations, and high-risk areas. The implications of this research extend to proactive decision-making for policymakers and transportation authorities. By providing accurate predictions and quantifiable insights into expected accident rates under different conditions, the paper aims to empower authorities to allocate resources efficiently and implement targeted interventions. The goal is to contribute to the development of informed policies and interventions that enhance road safety, creating a safer environment for all road users.

\end{abstract}


% \fi
\IEEEpeerreviewmaketitle

\section{Introduction}
The advent of motorized vehicular transportation has dramatically transformed global mobility, shaping economies, urban landscapes, and social dynamics. However, this rapid expansion of vehicle usage has also brought significant challenges, particularly in terms of road safety. The increase in the number of vehicles on the road has led to a corresponding rise in traffic accidents, resulting in substantial human, economic, and societal costs. As a result, predicting and preventing road accidents has become a critical area of research. In recent years, the application of machine learning techniques for predicting and analyzing road traffic accidents has gained significant attention. Numerous studies have explored various approaches to enhance road safety and predict accident occurrences.\\

The dataset used in this research was sourced from an online forum that records nationwide road accident incidents, including environmental factors and severity, as detailed in [7]. While the original dataset authors focused on classifying accident severity, our research repurposed the data to predict accident occurrences. The study in [1] analyzed and visualized the dataset, comparing classification models for accident severity. In contrast, our research conducts predictive analysis to forecast accident numbers at specific locations and times across the United States, moving beyond severity classification. This shift allows for a proactive approach to road safety, offering insights for targeted interventions and resource allocation. Our research builds on [2], where advanced transformer models predicted vehicle trajectories and accident likelihood. Unlike this vehicle-centric approach, we focused on environmental factors like weather, road types, and traffic patterns to improve prediction accuracy. By concentrating on these external factors, our model provides a broader understanding of the conditions leading to accidents, offering comprehensive predictive insights. This research aligns with broader efforts in machine learning for road safety, as seen in studies like [3], which introduced a fuzzy-RBF-based approach for accident prediction, and [4], which utilized deep learning and ensemble methods for fatality prediction. 

Our approach differs by focusing on predicting accident numbers rather than just categorizing severity, filling a critical gap in existing literature and offering practical applications for policy-making. Additionally, insights from this study could inform strategies similar to those in [5], where road safety prediction strategies were reviewed for IoV-enabled systems, showcasing the versatility of predictive analytics. Our methodology complements studies like [6], which analyzed machine learning algorithms for accident prediction, and [8], which used conflict coefficients for traffic accident prediction. Our research also aligns with techniques explored in [9] and [10], which determined traffic accidents using spatial-temporal techniques and aimed at reducing accident rates, respectively. Finally, our predictive approach is informed by methodologies in [11] and [12], which applied 3D scene reconstruction and analyzed environmental factors in traffic accidents. By integrating these varied methodologies, our research offers a comprehensive perspective on accident prediction. The use of data visualization and machine learning, as seen in [13] and [14], and the comparative analysis of algorithms in [15], further underscores the importance of our predictive model, which aims not only to classify but also to forecast accident occurrences, enabling more effective road safety interventions.\\

Our work stands out by focusing on predicting the actual number of accidents at specific locations and times across the United States rather than just classifying accident severity. We integrate environmental factors using Random Forest regression to enhance prediction accuracy. This shift from classification to prediction allows for more effective resource allocation and targeted interventions, making our approach more practical and impact in improving road safety compared to prior research. We utilized a Random Forest Regressor to predict the number of traffic accidents at specific locations and times across the United States. This machine learning technique was chosen for its ability to handle large datasets and its robustness in capturing complex patterns within the data. By leveraging the ensemble learning approach of Random Forest, we achieved high accuracy in forecasting accident occurrences, offering valuable insights for proactive road safety interventions.\\

\section{ Related Work}
With the help of  data science and machine learning models, the key achievements like different kinds of analysis, identifying the reason for accidents were achieved by the use of historical data. This section delves with the most latest works which is used to achieve the goal providing an overview of significant studies within the research community [1].\\

Sobhan Moosavi and team proposed the DAP model [7]. The maximum F1 score achieved by the model involving several neural network layers. While promising, the authors recommend to incorporate the latest methodologies for data cleaning and prepping steps which would significantly improve the accuracy. [7].\\

Honglei Ren and colleagues introduced a technique called LSTM, which models long-term dependencies in sequential data, for assessing hazards in traffic accidents. Their study, showing the prediction accuracy(RMSE) of 0.034, revealed the non-uniform distribution of traffic accidents in space and time.[8]. However, limitations were noted in the dataset scope, highlighting specific restrictions and temporal analysis. [9].\\

Lu Wenqi designed the TAP-CNN model, utilizing a Convolution Neural Network (CNN) architecture, achieving a prediction accuracy of 78.5\% [10]. The evaluation employed United States I-15 highway accident data within a restricted geographic area. Notably, the study focused on a small region, potentially limiting the generalizability of the model [11].\\
 
Senk and his team explored a model that predicts accidents by detecting hazardous areas through spatial pattern analysis of historical accident data [12]. But the model did not consider key factors such as time and weather, which could be major causes of accidents [13].\\

Tessa K. Anderson proposed k-means algorithm for hotspot profiling of accidents, based on data from the UK Police Department. Their research showed accident hotspots and patterns by mapping their density in various regions. The consideration of whether to develop region-specific or generalized systems for accident prediction is discussed [14, 15].

\section{ Methodology}
 \subsubsection{Dataset}
This research utilizes the dataset of accidents occurring in the United States, which is continuously updated with records obtained from the MapQuest and Bing APIs. The dataset snapshot used for this study covers the period from February 2016 to March 2023, comprising a substantial 7,728,394 rows and 47 columns of accident records.\\

The dataset contains various attributes that provide detailed information about accidents. Each entry is uniquely identified by an \texttt{ID} and is associated with a \texttt{Severity} level, indicating the degree of damage to the vehicle. The \texttt{Start\_Time} and \texttt{End\_Time} fields represent when the accident started and ended, respectively, in local time. The \texttt{Start\_Lat} and \texttt{Start\_Lng} provide the starting latitude and longitude coordinates of the accident, while \texttt{End\_Lat} and \texttt{End\_Lng} represent the end coordinates.

The \texttt{Distance} attribute captures the length of traffic affected by the accident. Address-related fields include \texttt{Number}, which refers to the street number in the address, \texttt{Street}, \texttt{Side} (relative street side), \texttt{City}, \texttt{Country}, \texttt{State}, and \texttt{Zipcode}. Additionally, the \texttt{Timezone} indicates the local time zone, and the \texttt{Airport\_Code} shows the closest airport near the accident location.

Weather data at the time of the accident is also provided, including the \texttt{Weather\_Timestamp}, temperature (\texttt{Temperature(F)}), \texttt{Wind\_Chill(F)}, \texttt{Humidity(\%)}, \texttt{Pressure(in)}, and \texttt{Visibility(mi)}. The wind's characteristics are recorded through \texttt{Wind\_Direction} and \texttt{Wind\_Speed(mph)}, while precipitation is measured with \texttt{Precipitation(in)}. The \texttt{Weather\_Condition} field captures the weather status (rain, snow, fog, etc.).

Other attributes highlight features of the surroundings, such as \texttt{Amenity} (presence of any nearby amenities), \texttt{Bump} (indicating the presence of speed bumps), \texttt{Crossing} (whether there's a pedestrian crossing), and \texttt{Give\_Way} (whether there is a give-way sign). Additional fields include \texttt{Junction}, \texttt{No\_Exit}, \texttt{Railway}, \texttt{Roundabout}, \texttt{Station}, \texttt{Stop}, and \texttt{Traffic\_Calming} to indicate nearby traffic-related features.

Signals and loops are marked by \texttt{Traffic\_Signal} and \texttt{Turning\_Loop}, respectively. Lastly, the dataset provides information about the time of day using fields like \texttt{Sunrise\_Sunset}, \texttt{Civil\_Twilight}, \texttt{Nautical\_Twilight}, and \texttt{Astronomical\_Twilight}. These represent different phases of daylight based on solar, civil, nautical, and astronomical references.\\

 \subsubsection{Exploratory Data Analysis}
The sheer size of the dataset considered resulted in very interesting insights and patterns about the accidents using various data mining techniques. Number of accidents per state: Fig. 1 demonstrates that California ranks highest in accident count, with Florida in second place.
\begin{figure}[ht]
\centering
\includegraphics[width=0.3\textwidth]{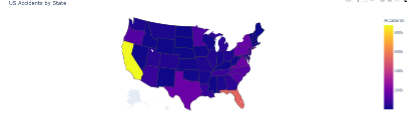}
\caption{US Accidents by State}
\label{fig:us_accidents}
\end{figure}

Analysis of Accident Occurrence Throughout the Day: Fig. 2 demonstrates the number of accidents occurring at different hours of the day. It is noticed that accidents are less likely to occur in the early morning, specifically from 4 to 6 am. Conversely, the highest likelihood of accidents is observed at the end of office hours, which spans from 5 to 7 pm.\\
\begin{figure}[ht]
\centering
\includegraphics[width=0.4\textwidth]{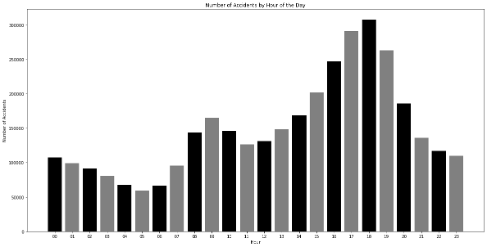}
\caption{Analysis of Accident Occurrence Throughout the Day}
\label{fig:accident_analysis}
\end{figure}

Daily, Monthly and yearly accident Frequency: Fig.3, Fig. 4 and Fig. 5 demonstrates the daily, monthly and yearly accident frequency respectively. It is noticed that the accident count sharply increased in 2019 and doubled in 2020. In the graph number of accidents per month it can be noticed that the cold months have a greater number of accidents as compared to the other months. It is noted that there is an increase in number of accidents on 22 and 23.\\

\begin{figure}[ht]
\centering
\includegraphics[width=\linewidth]{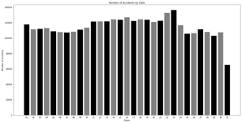}
\caption{Daily accident frequency}
\end{figure}
\begin{figure}[ht]
\centering
\includegraphics[width=\linewidth]{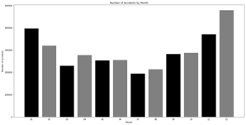}
\caption{Monthly accident frequency}
\end{figure}
\begin{figure}[ht]
\centering
\includegraphics[width=\linewidth]{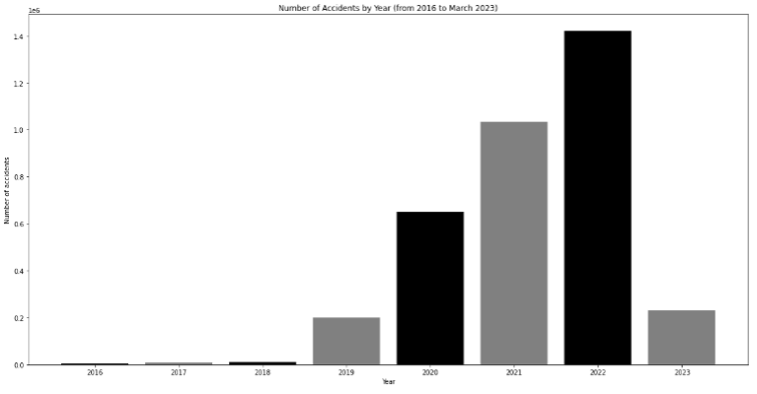}
\caption{Yearly accident frequency}
\end{figure}

Weekly accident frequency: Fig. 6 shows a bar graph displaying the distribution of accidents across different days of the week. Friday appears to have the highest proportion of accidents at 17.2\%,  closely followed by Thursday at 16.1\%. Wednesday also sees a significant number of incidents, accounting for 15.7\% of the total. Interestingly, Sunday, often considered a rest day, records a relatively lower percentage of accidents at 9.7\%. The other days, namely Monday, Tuesday, and Saturday, show comparable percentages, hovering around 11\% to 15\%. The distribution suggests that while accidents are a consistent concern throughout the week, certain days, like Fridays and Thursdays, experience a slightly elevated number of incidents. This information could be vital for traffic management and safety advisories.\\
\begin{figure}[ht]
\centering
\includegraphics[width=\linewidth]{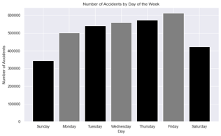}
\caption{Weekly accident frequency}
\end{figure}

Number of accidents versus Windspeed: Fig. 7 illustrates the relationship between wind speed (measured in mph on the horizontal axis), and the number of accidents (depicted on the vertical axis) indicating higher wind speeds may not directly correlate with a surge in accidents. It is evident that the highest number of accidents occurs at very low wind speeds, close to 0 mph. As the wind speed increases, particularly around 5 to 15 mph, the number of accidents experiences notable fluctuations, showing several peaks and troughs. Beyond 20 mph, the number of accidents stabilizes and remains relatively low.\\
\begin{figure}[ht]
\centering
\includegraphics[width=\linewidth]{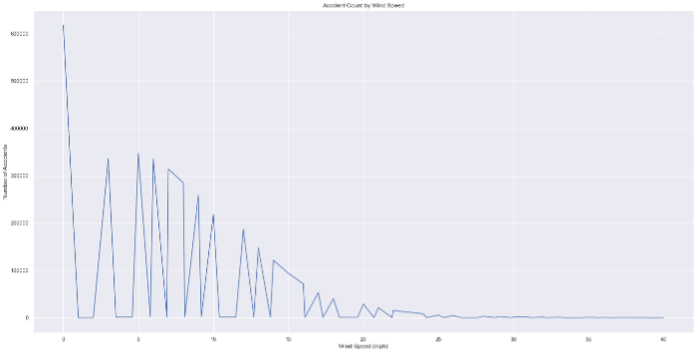}
\caption{Number of accidents versus Wind-speed}
\end{figure}
\section{Experiments and Results}
The Random Forest Regressor emerges as the top model with an MSE at 85.21 and R-Square score at 0.75, indicating a strong fit for the data. The predicted value is 1.33, closely matching the actual value of 1, highlighting its accuracy and reliability in delivering precise predictions. This performance underscores the effectiveness of the Random Forest Regressor in capturing complex data patterns and providing robust predictive power. The table also includes a noteworthy entry for Auto Machine Learning, which demonstrates competitive performance with an MSE of 72.32 and an R-Square of 0.64. The predicted value of 1.84 suggests that automated machine learning approaches can effectively optimize model selection and hyperparameter tuning, making them a valuable addition to the modeling toolkit.\\

The LightGBM (LGBM) Regressor also shows commendable performance with an MSE of 108.01 and an R-Square of 0.68, producing a predicted value of 2.99. This model's performance further supports the utility of gradient boosting techniques in providing reliable predictions. The Decision Tree Regressor maintains moderate performance with an MSE of 145.41 and an R-Square of 0.57, producing a predicted value of 1.47. While not as strong as the Random Forest or LightGBM Regressors, the Decision Tree Regressor offers a balance between simplicity and predictive accuracy, making it a useful model for certain applications.\\

Conversely, Adaptive Boosting, Lasso Regression, Elastic Net Regression, and Robust Regression exhibit relatively high MSE values and low R-Square values, indicating poorer fits to the data. For instance, Adaptive Boosting has an MSE of 327.22 and an R-Square of 0.04, with a predicted value of 7.84, significantly deviating from the actual value. Similarly, Lasso Regression and Elastic Net Regression have MSE values of 335.41 and 334.77, respectively, both with R-Square values of 0.01 and 0.02, predicting values that are far from the actual. The LSTM model, typically utilized for sequence prediction tasks, shows moderate performance with an MSE of 153.97 and an R-Square of 0.55, predicting a value of 16.88. This indicates that while LSTM can capture temporal dependencies, it may not be the best choice for this specific regression task.\\

The Category Boosting model performs moderately well with an MSE of 116.60 and an R-Square of 0.63, predicting a value of 8.15. This performance is better than some other models but still shows room for improvement. 

The Stacking Regressor displays strong performance with an MSE of 52.08 and an R-Square of 0.73. Its predicted value of 1.71 is reasonably close to the actual value, demonstrating its effectiveness in combining the strengths of multiple models to enhance predictive accuracy.

\begin{figure}[ht]
\centering
\includegraphics[width=\linewidth]{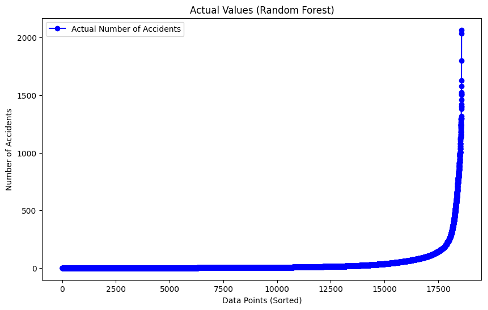}
\caption{Actual values in Random Forest}
\end{figure}
\begin{figure}[ht]
\centering
\includegraphics[width=\linewidth]{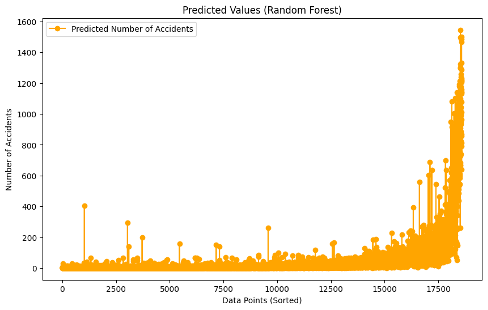}
\caption{Predicted values in Random Forest}
\end{figure}

\begin{table}
\centering

\begin{tabular}{| l  |l  |l  |l  |l |}
\hline
\textbf{Models}& \textbf{MSE}&  \textbf{R-Square}& \textbf{Actual}& \textbf{Predicted}\\
\hline
Random Forest Regressor & 85.21 & 0.75& 1 & 1.33 \\
\hline
Auto Machine Learning & 72.32 & 0.64& 1 & 1.84 \\
\hline
Decision Tree Regressor & 145.41& 0.57& 1 & 1 \\
\hline
Adaptive Boosting & 327.22 & 0.04& 1 & 7.84\\
\hline
Category Boosting & 116.50 & 0.65& 1 & 8.15\\
\hline
Long-Short Term Memory & 153.97 & 0.55& 1 & 16.88\\
\hline
Lasso Regression & 335.41 & 0.01& 1 & 8.59 \\
\hline
Elastic Net Regression & 334.74 & 0.02& 1 & 8.80\\
\hline
Robust Regression & 352.56& -0.03& 1 & 4.08\\
\hline
Stacking Regressor  & 82.08& 0.75& 1 & 1.71\\
\hline
LGBM Regressor & 108.01& 0.68& 1 & 2.99\\
\hline

\end{tabular}
\vspace{0.01cm} % Adjust the space as needed
\caption{Results comparing various algorithms}
\label{tab:my_table}
\end{table}

\section{Conclusion}
% The conclusion goes here.

This research provides a comprehensive analysis of traffic accident prediction across the United States, leveraging an extensive dataset encompassing 49 states. By applying machine learning models, particularly the Random Forest Regressor, we were able to accurately predict accident occurrences and identify high-risk areas. The findings underscore the potential of predictive analytics to enhance road safety by enabling proactive interventions based on data-driven insights.

The Random Forest Regressor demonstrated the highest predictive accuracy, making it an effective tool for understanding the complex factors contributing to traffic accidents at a national scale. The use of this model has allowed us to capture significant patterns and trends, providing valuable insights that can inform the development of targeted policies and strategies aimed at reducing accident rates and improving emergency response.

This research not only fills a critical knowledge gap but also lays the groundwork for evidence-based decision-making in public safety. By identifying key risk factors and accident hotspots, policymakers and law enforcement agencies can prioritize resource allocation, optimize safety measures, and ultimately reduce the occurrence and severity of traffic accidents across the nation.


% \section*{Acknowledgment}




%\bibliographystyle{IEEEtran}

\end{document}